\documentclass[twocolumn,showpacs,aps,prl,superscriptaddress,letterpage]{revtex4}

\usepackage{graphicx}
\usepackage{dcolumn}
\usepackage{amsmath}
\usepackage{epsfig}

\input babarsym

\newcommand{\BABARPubYear}    {11}
\newcommand{\BABARPubNumber}  {012}

\newcommand{\SLACPubNumber} {14465}

\def\figurebox#1#2#3{%
    \def\arg{#3}%
    \ifx\arg\empty
    {\hfill\vbox{\hsize#2\hrule\hbox to #2{\vrule\hfill\vbox to #1{\hsize#2\vfill}\vrule}\hrule}\hfill}%
    \else
    {\hfill\epsfbox{#3}\hfill}%
    \fi}

\begin{document}
\preprint{\babar-PUB-\BABARPubYear/\BABARPubNumber} 
\preprint{SLAC-PUB-\SLACPubNumber} 

\begin{flushleft}
\babar-PUB-\BABARPubYear/\BABARPubNumber\\
SLAC-PUB-\SLACPubNumber\\
\end{flushleft}

\title{
{\large \bf
\boldmath
Observation of the baryonic \B decay $\Bzb \ra \LCp \Lbar \Km$
\unboldmath
} 
}

%
\author{J.~P.~Lees}
\author{V.~Poireau}
\author{V.~Tisserand}
\affiliation{Laboratoire d'Annecy-le-Vieux de Physique des Particules (LAPP), Universit\'e de Savoie, CNRS/IN2P3,  F-74941 Annecy-Le-Vieux, France}
\author{J.~Garra~Tico}
\author{E.~Grauges}
\affiliation{Universitat de Barcelona, Facultat de Fisica, Departament ECM, E-08028 Barcelona, Spain }
\author{M.~Martinelli$^{ab}$}
\author{D.~A.~Milanes$^{a}$}
\author{A.~Palano$^{ab}$ }
\author{M.~Pappagallo$^{ab}$ }
\affiliation{INFN Sezione di Bari$^{a}$; Dipartimento di Fisica, Universit\`a di Bari$^{b}$, I-70126 Bari, Italy }
\author{G.~Eigen}
\author{B.~Stugu}
\author{L.~Sun}
\affiliation{University of Bergen, Institute of Physics, N-5007 Bergen, Norway }
% ...
\collaboration{The \babar\ Collaboration}
\noaffiliation

\date{August 16, 2011}

\begin{abstract}
We report the observation of the baryonic \B decay $\Bzb \ra \LCp \Lbar \Km$ with a significance larger than $7$ standard deviations based on $471 \times 10^6$ \BBb pairs collected with the \babar~detector at the \pep2 storage ring at SLAC.
We measure the branching fraction for the decay $\Bzb \ra \LCp \Lbar \Km$ to be $(3.8 \pm 0.8_{\rm stat} \pm 0.2_{\rm sys} \pm 1.0_{\LCp}) \times 10^{-5}$. 
The uncertainties are statistical, systematic, and due to the uncertainty in the $\LCp$ branching fraction. 
We find that the $\LCp \Km$ invariant mass distribution shows an enhancement above $3.5\gevcc$.
\end{abstract}

\pacs{13.25.Hw, 13.60.Rj, 14.20.Lq}

\maketitle
While baryons are produced in $(6.8 \pm 0.6)\%$ \cite{ref:PDG} of all \B-meson decays, little is known about the detailed mechanics of these decays and more generally about hadron fragmentation into baryons. 
We can increase our understanding of baryon production in \B decays by comparing decay rates for related exclusive final states.
In this paper we present a measurement of the decay $\Bzb \ra \LCp \Lbar \Km$ \cite{footnote}. 
Currently, no experimental results are available for this decay.

This analysis is based on a dataset of about $429 \invfb$, corresponding to $471 \times 10^6$ \BBb pairs, collected with the \babar detector at the \pep2 asymmetric-energy \epem storage ring, operated at a center-of-mass energy equal to the \FourS mass. 
The signal efficiency is determined with a Monte Carlo simulation based on EvtGen \cite{ref:EvtGen} for the event generation, and GEANT4 \cite{ref:geant} for the detector simulation. The $\Bzb \ra \LCp \Lbar \Km$ Monte Carlo events are generated uniformly in the $\LCp \Lbar \Km$ phase space. Monte Carlo simulated events are used to study background contributions as well.

The \babar detector is described in detail elsewhere \cite{ref:NIM}. 
Charged particle trajectories are measured by a five-layer double-sided silicon vertex tracker and a 40-layer drift chamber, both immersed in a $1.5$ T axial magnetic field. Charged particle identification is provided by ionization energy measurements along with Cherenkov radiation detection by an internally reflecting ring-imaging detector (DIRC).

The \LCp is reconstructed in the decay mode $\LCp \ra \proton \Km \pip$ and the $\Lbar$ in the decay mode $\Lbar \ra \antiproton \pip$. 
For the identification of proton, kaon, and pion candidates, we use selection criteria based on the measurements of the specific ionization in the tracking detectors, and of the Cherenkov radiation in the DIRC \cite{Aubert:Btohh}.

For the identification of the \proton coming from the \LCp the average efficiency is about $95\%$ while the probability of misidentifying a kaon as a proton is less than $2\%$. The average efficiency for the \Km identification is about $90\%$
The probability of misidentifying a pion as a kaon is about $5\%$.
These are the dominant misidentification probabilities for each particle type.
The \LCp daughters and the $\Lbar$ daughters are each fit to a common vertex and the \LCp and the $\Lbar$ candidate invariant mass is required to lie within $3 \sigma$ of the world average mass\cite{ref:PDG}; i.e., in the range $2.273$ to $2.299 \gevcc$ and $1.113$ to $1.119 \gevcc$, respectively.
For the reconstruction of the \B candidate, the mass of the $\LCp$ candidate is constrained to its nominal value \cite{ref:PDG} and is combined with a $\Lbar$ and a \Km candidate. 
Since the $\Lbar$ candidate mass is already well measured, it is not constrained.

The \LCp, $\Lbar$ and \Km candidates are then fitted to a common vertex and the confidence level of this fit is required to exceed $0.2\%$.

A possible source for fake signal events is the decay $\Bzb \ra \LCp \antiproton \Km \pip$ \cite{ref:2009wn}, which has the same final state as the decay under investigation. In order to suppress this background we require that the distance between the \B vertex and the $\Lbar$ vertex in the $xy$ plane (with $z$ parallel to the beam axis) exceeds $0.4\cm$. 
This constraint reduces combinatoric background by $18\%$, and the background from $\Bzb \ra \LCp \antiproton \Km \pip$ by $99.6\%$. The expected remaining background from this decay is determined to be $0.1 \pm 0.1$ events\cite{ref:2009wn}.

The separation of signal and background in the candidate sample is obtained by using two kinematic variables, $\DeltaE = E_{\B}^* - \sqrt{s}/2$ and $\mes = \sqrt{(s/2+\mathbf{p}_i \cdot \mathbf{p}_{\B})^2/E_i^2-\left|\mathbf{p}\right|_{\B}^2}$, where $\sqrt{s}$ is the $\epem$ center-of-mass energy and $E_{\B}^*$ the energy of the \B candidate in the center-of-mass system. $(E_i, \mathbf{p}_i)$ is the four-momentum vector of the \epem system and $\mathbf{p}_{\B}$ the \B-candidate momentum vector, both measured in the laboratory frame.
For true \B decays \mes is centered at the \B-meson mass and \DeltaE is centered at zero. \B candidates are required to have an \mes value between $5.272$ and $5.288 \gevcc$.

\begin{figure}[h]
	\centering\includegraphics[width=.48\textwidth]{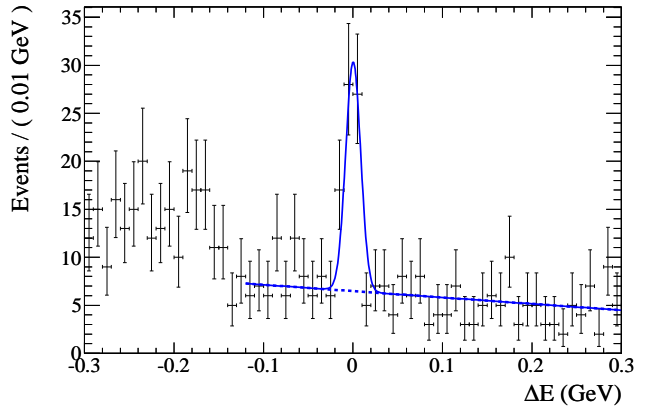}
	\caption{The \DeltaE distribution for $\LCp \Lbar \Km$ candidates in data with all selection criteria applied (points). The solid line is the overall fit result, while the dashed line is the background component.}
	\label{fig:DeltaE}
\end{figure}

Figure~\ref{fig:DeltaE} shows the \DeltaE distribution of the selected candidates, fitted in the range from $-0.12$ to $0.30 \gev$. 
We fit the signal with a Gaussian with the mean $\mu$ and width $\sigma$ fixed to the values obtained from a fit to the Monte Carlo simulation ($\mu = 0.247\mev$ and $\sigma = 8.381\mev$), leaving only the signal yield floating.
The background is described by a first-order polynomial.
A binned maximum likelihood fit with this probability density function (PDF) gives a signal yield of $51 \pm 9$ events. (For the branching fraction measurement described later, we use the excess number of candidates above background as the estimate of the number of signal events.)
The confidence level for the null hypothesis, considering statistical uncertainties only, is $2.6\times 10^{-15}$, which corresponds to a statistical significance of $8$ standard deviations.
A possible background from $\Bzb \ra \LCp \bar{\Sigma}^0 \Km$, which rises slowly up to $\DeltaE \approx -0.06\gev$ and drops sharply between $\DeltaE = -0.05\gev$ and $-0.02\gev$, is not visible in Fig.~\ref{fig:DeltaE}.

\begin{figure}[h]
	\centering\includegraphics[width=.48\textwidth]{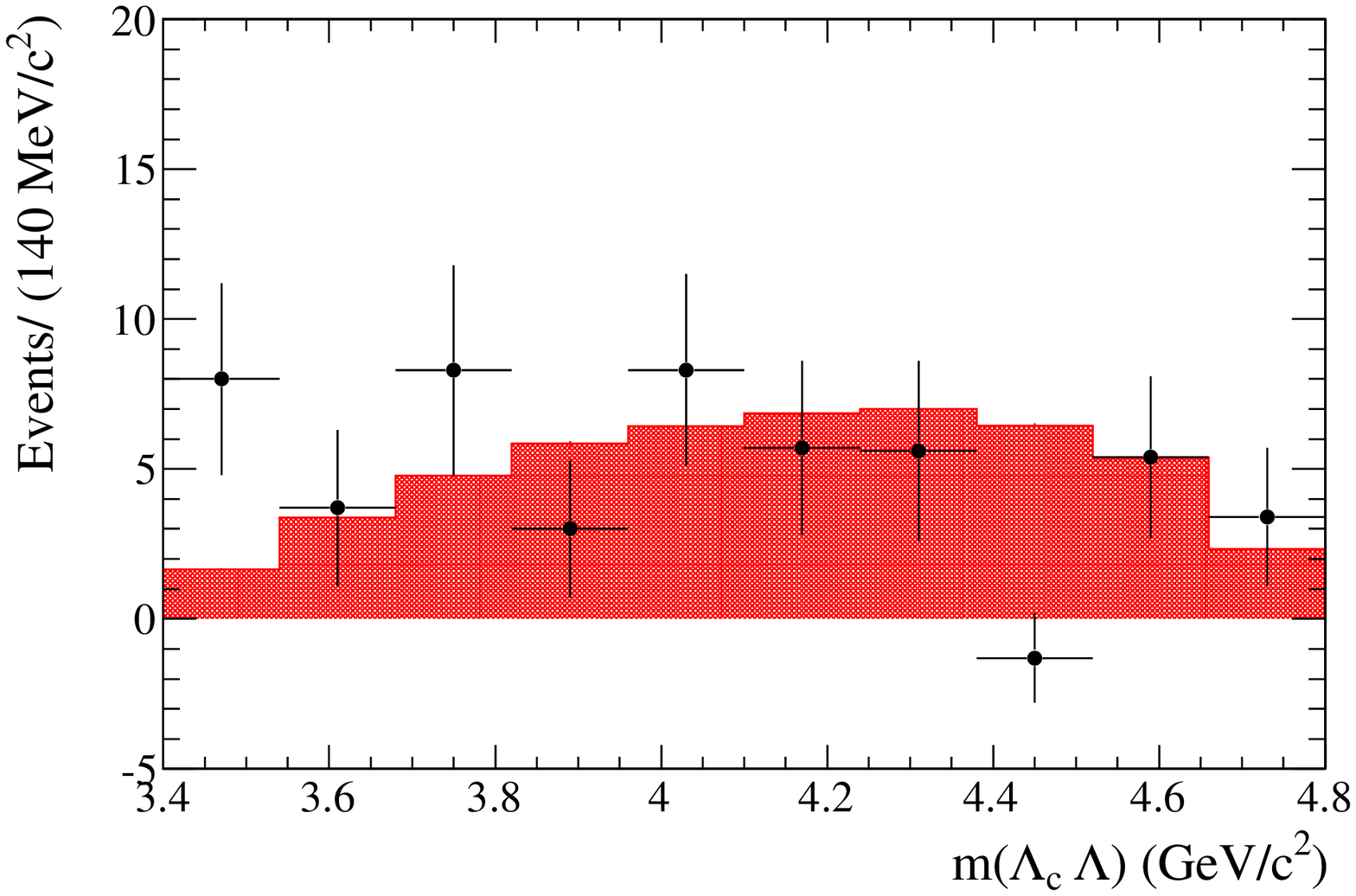}
	\centering\includegraphics[width=.48\textwidth]{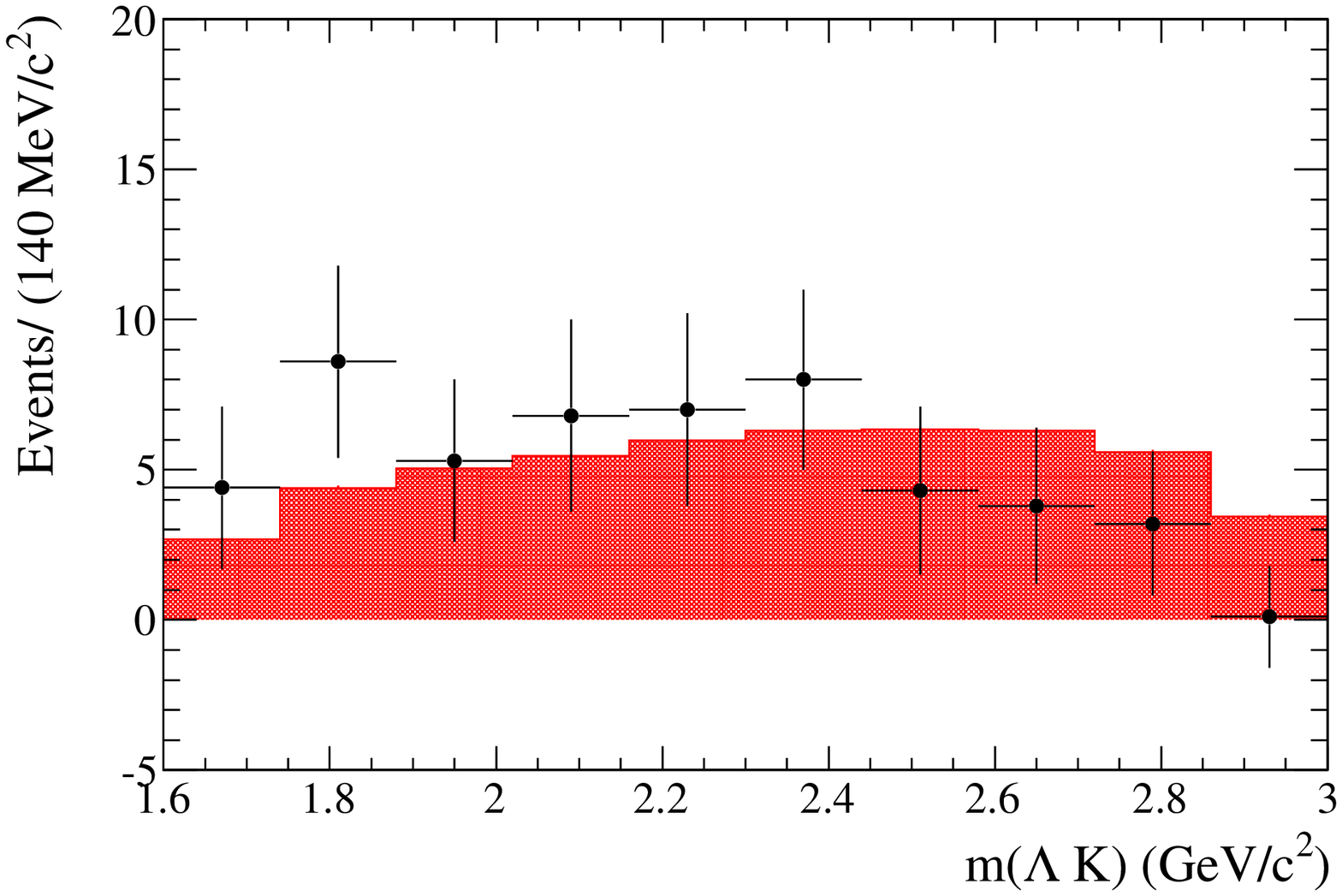}
	\centering\includegraphics[width=.48\textwidth]{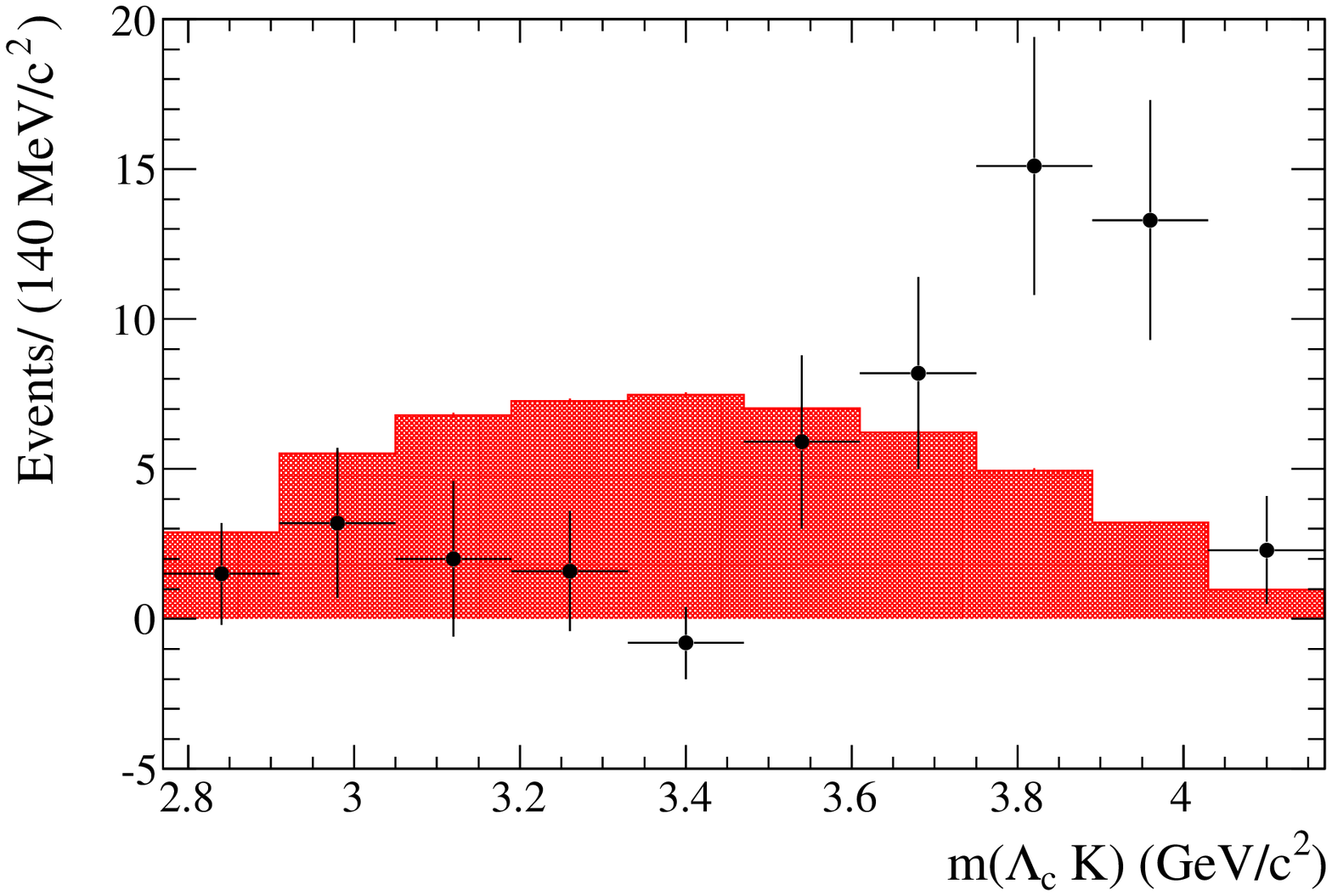}
	\caption{The $m(\LCp \Lbar)$, $m(\Lbar \Km)$ and $m(\LCp \Km)$ distributions in data (points) in comparison with the Monte Carlo sample (red histogram).}
	\label{fig:Compare_orig}
\end{figure}

Since the decay dynamics of baryonic \B decays are largely unknown, we investigate the invariant-mass distribution of the two-body systems. 
Intermediate states would appear as differences in the invariant-mass distribution for date and for the $\Bzb \ra \LCp \Lbar \Km$ Monte Carlo simulation, in which the final state is generated according to three-body phase space.
Using the same function that we used in Fig.~\ref{fig:DeltaE}, we fit the \DeltaE distributions for ten ranges of the three two-body masses.
The results are compared to the phase space Monte Carlo simulation in Fig.~\ref{fig:Compare_orig}. 
While the $m(\LCp \Lbar)$ and $m(\Lbar \Km)$ distributions show no significant deviations, the $m(\LCp \Km)$ distribution shows the data concentrated in the upper half of the allowed mass range, contrary to the Monte Carlo simulation. 
A possible explanation for this is a resonant decay via a baryon resonance that has not yet been observed. Another possibilty is enhanced rates at both $m(\LCp \Lbar)$ and $m(\Lbar \Km)$ thresholds.

\begin{figure}[h]
	\centering\includegraphics[width=.48\textwidth]{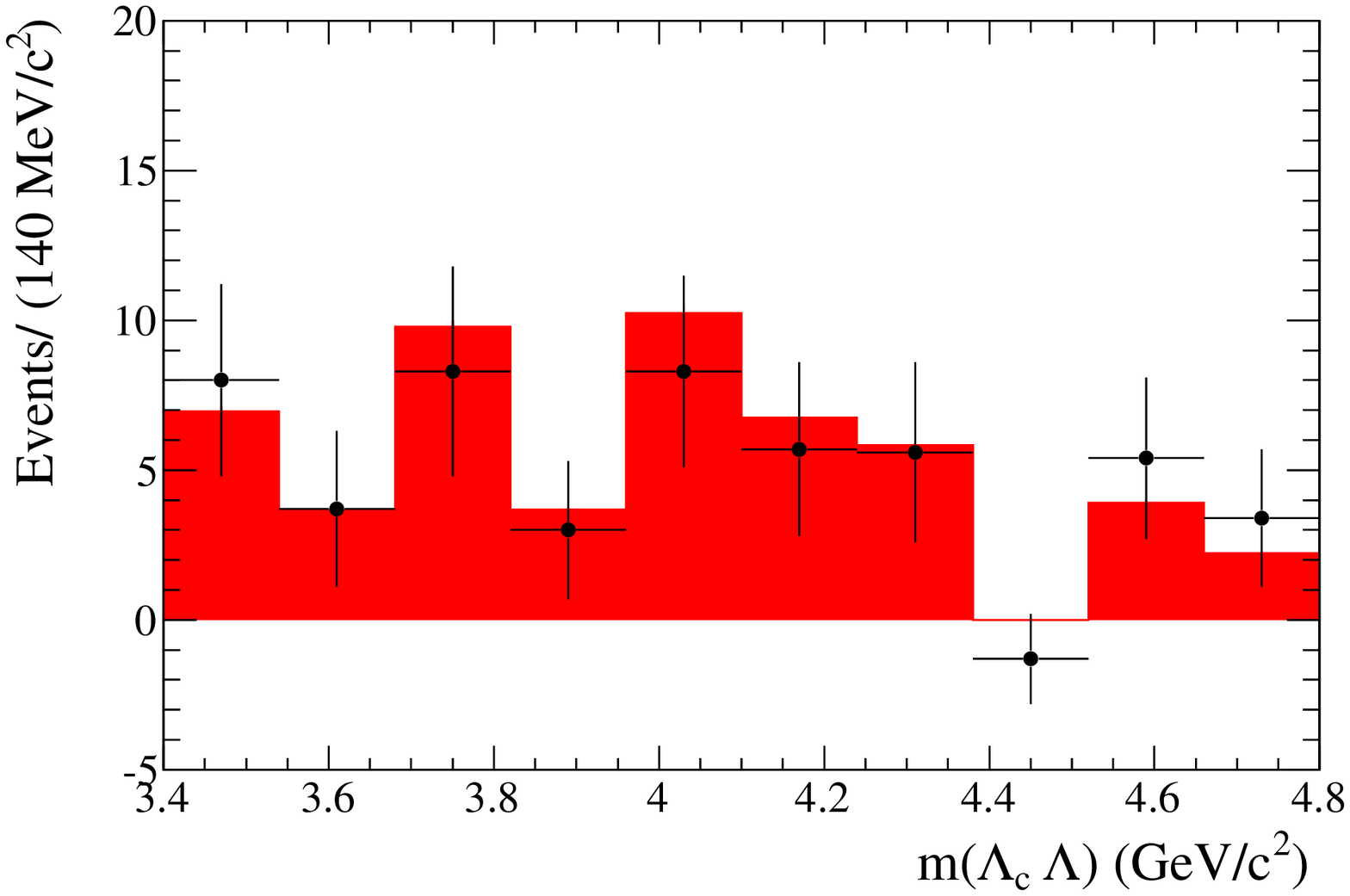}
	\centering\includegraphics[width=.48\textwidth]{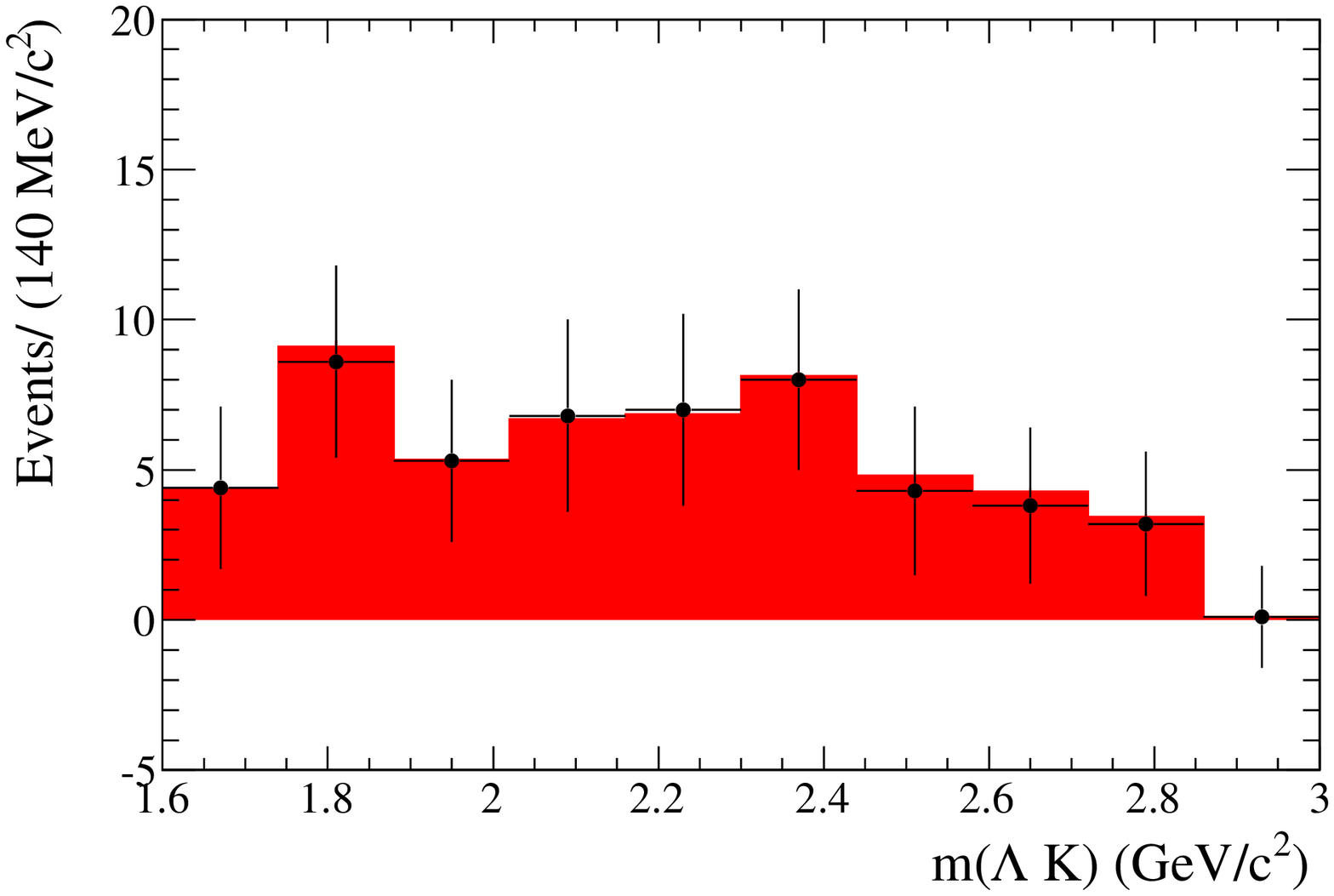}
	\centering\includegraphics[width=.48\textwidth]{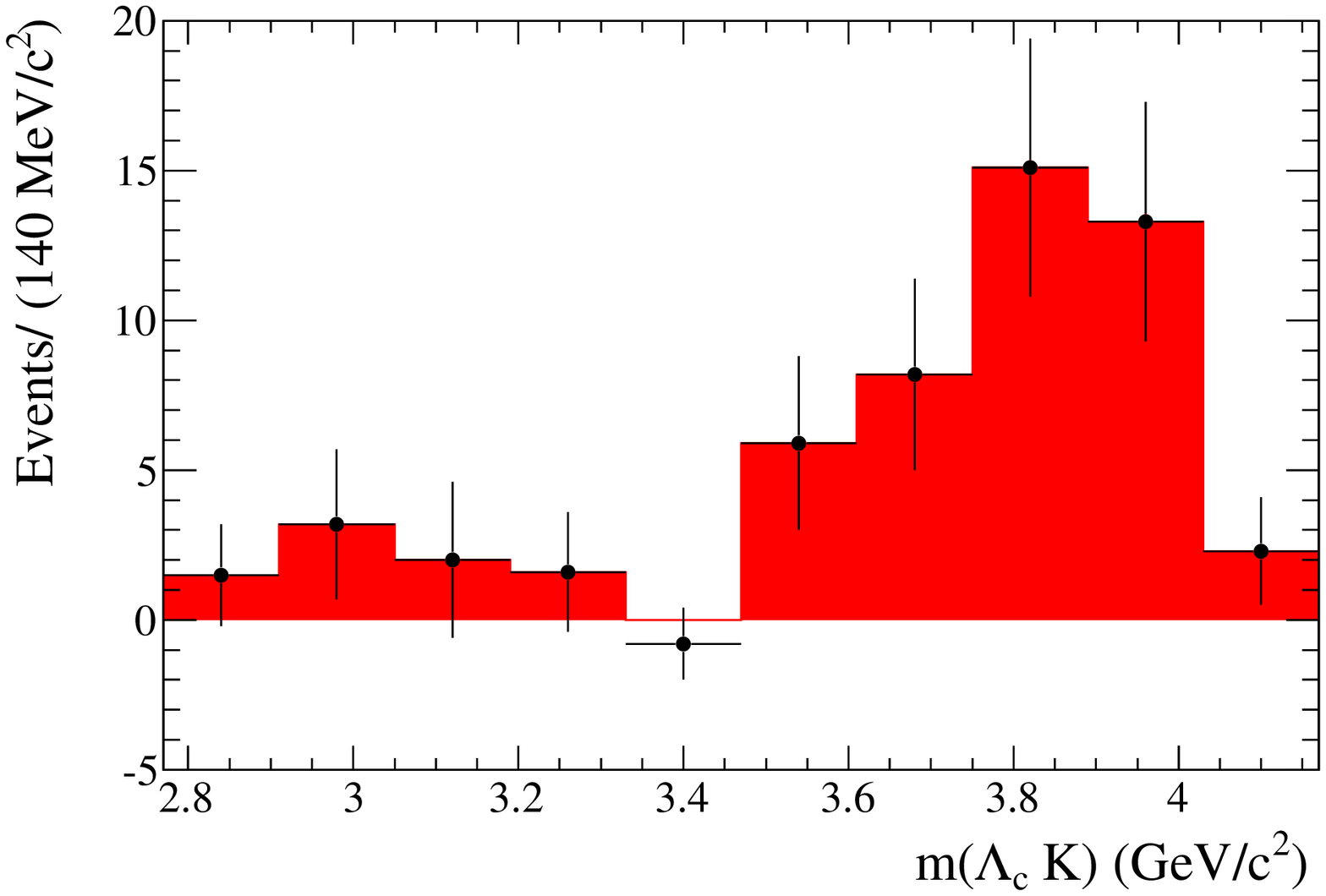}
	\caption{The background-subtracted $m(\LCp \Lbar)$, $m(\Lbar \Km)$ and $m(\LCp \Km)$ distributions for data (points) and for the final weighted Monte Carlo sample (red histogram).}
	\label{fig:Compare_final}
\end{figure}

Because the efficiency varies over the Dalitz plot and the distribution of candidates in data is unknown {\em a priori}, we must use the distribution of the data events in the Dalitz plot to estimate the efficiency. The small number of candidates, combined with resolution and edge effects, make the simple weighting of events by the inverse of the efficiency problematic. Instead we determine a set of weights to apply to the simulated events so that the resulting weighted Monte Carlo distributions mimic the data.
We make the assumption that the dependence of the decay dynamics on the two-body invariant masses can be factorized into the product of three functions that each depend on one invariant mass and weight the Monte Carlo events with the function:
\begin{align}
	w&[m(\LCp \Km),m(\LCp \Lbar), m(\Lbar \Km] \notag\\
	&= w_a[m(\LCp \Km)] \cdot w_b[m(\LCp \Lbar)] \cdot w_c[m(\Lbar \Km)].
	\label{eq:1}	
\end{align}
By dividing the background subtracted $m(\LCp \Km)$ distribution (Fig.~\ref{fig:Compare_orig}) by the corresponding distribution from the phase-space Monte Carlo simulation, we obtain the weights $w_a[m(\LCp \Km)]$, which are used to weight the Monte Carlo candidates. (If a negative weight is required the weight is constrained to zero.)
Next, we use the weighted Monte Carlo candidates to determine $w_b[m(\LCp \Lbar)]$ in the same way and then use $w_a[m(\LCp \Km)] \cdot w_b[m(\LCp \Lbar)]$ to determine $w_c[m(\Lbar \Km)]$.
After each weighting we determine the reconstruction efficiency by a fit to the \DeltaE distribution for weighted Monte Carlo.
Starting with these weights, the weighting is repeated until the reconstruction efficiency converges and the two-body mass distributions in data and Monte Carlo agree within statistical uncertainties. The efficiency after each weighting is shown in Table~\ref{tab:weigh_results}.
\begin{table}
	\caption{Two-body invariant mass distributions used for the weighting and the corresponding reconstruction efficiency $\varepsilon$. The dots indicate how often the respective mass distributions are used to determine the weights.}
	\begin{center}
		\begin{tabular}{cccc}\hline\hline\\[-2ex]
			$m(\LCp \Km)$\hspace{.2cm}	& $m(\LCp \Lbar)$\hspace{.2cm}	& $m(\Lbar \Km)$\hspace{.2cm}	& Efficiency $\varepsilon$ in $\%$	\\\hline
							&				&				& $10.90$		\\
			$\bullet$			&				&				& $8.60$		\\
			$\bullet$			& $\bullet$			&				& $9.21$		\\
			$\bullet$			& $\bullet$			& $\bullet$			& $9.19$		\\
			$\bullet$$\bullet$		& $\bullet$			& $\bullet$			& $8.81$		\\
			$\bullet \bullet$		& $\bullet \bullet$		& $\bullet$			& $8.80$		\\
			$\bullet \bullet$		& $\bullet \bullet$		& $\bullet \bullet$		& $8.77$		\\
			\hline\hline
		\end{tabular}
	\end{center}
	\label{tab:weigh_results}
\end{table}
Since the $m(\LCp \Km)$ distribution in data shows the strongest deviations compared to the phase space Monte Carlo simulation we use the efficiency $\varepsilon$ obtained after the second weighting in $m(\LCp \Km)$ [$\varepsilon = 8.81\%$]. The comparison between data and weighted Monte Carlo events in the two-body masses can be seen in Fig.~\ref{fig:Compare_final}. Note that, by construction, the data and simulation agree exactly for the $m(\LCp \Km)$ distribution. The close agreement in the other two distributions shows that the form given in Eq.~\eqref{eq:1} is adequate to describe any correlations between variables in the data. 
The effect of the statistical uncertainties in the data on the efficiency determination is described below.

For the branching fraction calculation we determine the number of reconstructed events by a fit to \DeltaE with a first-order polynomial for the background. To avoid a potential bias introduced by an assumption of the signal shape, we fit the region $-0.12 < \DeltaE < 0.30 \gev$ exclusive of the signal region $-0.03 < \DeltaE < 0.03\gev$. By extrapolating the background yield into the signal region and subtracting it from the integral of the histogram in this region we obtain a signal yield of $N_{\rm sig} = 50 \pm 11$. This results in a branching fraction of
\begin{align}
	\BR&(\Bzb \ra \LCp \Lbar \Km) \notag\\
	 &= \frac{N_{\rm sig}/\varepsilon}{N_{\BBb} \cdot \BR(\LCp \ra \proton \Km \pip) \cdot \BR(\Lbar \ra \antiproton \pip)} \notag\\
	 &= (3.8 \pm 0.8_{\rm stat} \pm 1.0_{\LCp})\times 10^{-5},
\end{align}
with $N_{\BBb} = N_{\Bzb} + N_{\Bz} = (471 \pm 3) \times 10^{6}$, assuming equal production of $\BzBzb$ and $\BpBm$ in the decay of the \FourS. The branching fractions $\BR(\LCp \ra \proton \Km \pip) = (5.0 \pm 1.3)\%$ and $\BR(\Lbar \ra \antiproton \pip) = (63.9 \pm 0.5)\%$ are the world averages from Ref.~\cite{ref:PDG}.

Several sources of systematic uncertainties are investigated and summarized in Table~\ref{tab:sys}. Most of the uncertainties are derived from comparisons between Monte Carlo simulations and control samples in data. Systematic uncertainties arise from uncertainties in charged particle reconstruction efficiencies ($0.9\%$) and charged particle identification efficiencies ($2.4\%$), and from statistical uncertainties in the Monte Carlo simulation ($0.5\%$). The systematic uncertainty on the number of \BBb pairs is $0.6\%$.
The systematic uncertainty from the $\Lbar$ branching fraction amounts to $0.8\%$. 

The systematic uncertainty introduced by neglecting a possible $\Bzb \ra \LCp \bar{\Sigma}^0 \Km$ background is determined by adding a PDF for this background to the fit function used for the \DeltaE fit shown in Fig.~\ref{fig:DeltaE}. Allowing nonnegative contributions from this background, the fit returns a value of $0.0^{+ 1.8}_{- 0.0}$. For a conservative limit on this systematic uncertainty we fix the yield to $1.8$ and take the change in the number of signal events as systematic uncertainty ($1.0\%$). 
The \DeltaE distribution in Fig.~\ref{fig:DeltaE} shows an enhancement below $-0.14\gev$, caused by decays of the type $\B \ra \LCp \Lbar \Km \pi$. Due to the limited resolution these events could leak into the fit region from $-0.12$ to $0.30\gev$. We determine the resulting systematic uncertainty by changing the fit region to $-0.10$ to $0.30\gev$. The branching fraction changes by $1.8\%$.

The uncertainty arising from the chosen background description is determined by repeating the fit to determine the signal yield with a second-order polynomial for the background. 
The number of signal events changes by $2.0\%$.
A comparison between data and Monte Carlo events shows that the mean of the \DeltaE distribution in data is shifted by $-0.003\gev$. We determine the resulting systematic uncertainty by shifting the signal region for the fit in the Monte Carlo \DeltaE distribution by $0.003\gev$. This changes the efficiency to $8.60\%$, corresponding to a systematic uncertainty of $0.8\%$.

The uncertainty due to the treatment of the three-body phase space in the efficiency correction is estimated from the variation of the efficiency when performing further iterations of weighting. Table~\ref{tab:weigh_results} shows a variation from $8.81\%$ down to $8.77\%$. This corresponds to a systematic uncertainty of $0.5\%$. 

A possible additional contribution to the statistical uncertainty coming from the efficiency determination, where we determined the weights based on data events in ranges of the two-body masses, is studied by performing the efficiency correction in ranges of $m(\Lambda_c^+ K^-)$ only, with unweighted Monte Carlo events. The change in overall reconstruction efficiency is negligible compared to the statistical uncertainty.

Adding all contributions in quadrature we obtain a systematic uncertainty of $4.1\%$.
\begin{table}
	\caption{Summary of the relative systeamtic uncertainties on the branching fraction $\BR(\Bzb \ra \LCp \Lbar \Km)$.}
	\begin{center}
		\begin{tabular}{lc}\hline\hline
			Source					& Relative syst. uncert.		\\\hline
			Track reconstruction			& $0.9\%$				\\
			Charged particle ID			& $2.4\%$				\\
			$N_{\BBb}$				& $0.6\%$				\\
			$\BR(\Lbar \ra \antiproton \pip)$	& $0.8\%$				\\
			Monte Carlo statistics			& $0.5\%$				\\
			$\Bzb \ra \LCp \bar{\Sigma}^0 \Km$	& $1.0\%$				\\
			$\B \ra \LCp \Lbar \Km \pi$		& $1.8\%$				\\
			\DeltaE background description		& $2.0\%$				\\
			$\mu(\DeltaE)$ shift			& $0.8\%$				\\
			Efficiency determination		& $0.5\%$				\\\hline
			Total					& $4.1\%$				\\
			\hline\hline
		\end{tabular}
	\end{center}
	\label{tab:sys}
\end{table}
The significance of the signal, including additive systematic uncertainties is determined to be more than $7$ standard deviations. This significance includes systematic uncertainties from the $\Bzb \ra \LCp \bar{\Sigma}^0 \Km$ and $\B \ra \LCp \Lbar \Km \pi$ background as well as the \DeltaE background description.

In summary, we observe the decay $\Bzb \ra \LCp \Lbar \Km$ with a significance larger than $7$ standard deviations and measure a branching fraction of
\begin{align}
	\BR&(\Bzb \ra \LCp \Lbar \Km) \notag\\
	&= (3.8 \pm 0.8_{\rm stat} \pm 0.2_{\rm sys} \pm 1.0_{\LCp})\times 10^{-5}.
\end{align} 
The decay rate is not uniform over three-body phase space; rather, it is dominant at high $\LCp \Km$ mass.

We are grateful for the excellent luminosity and machine conditions
provided by our \pep2\ colleagues, 
and for the substantial dedicated effort from
the computing organizations that support \babar.
The collaborating institutions wish to thank 
SLAC for its support and kind hospitality. 
This work is supported by
DOE
and NSF (USA),
NSERC (Canada),
CEA and
CNRS-IN2P3
(France),
BMBF and DFG
(Germany),
INFN (Italy),
FOM (The Netherlands),
NFR (Norway),
MES (Russia),
MICIIN (Spain),
STFC (United Kingdom). 
Individuals have received support from the
Marie Curie EIF (European Union),
the A.~P.~Sloan Foundation (USA)
and the Binational Science Foundation (USA-Israel).


\begin{thebibliography}{99}
\bibitem{footnote}
Throughout this paper, all decay modes represent that mode and its charge conjugate.
\end{thebibliography}
\end{document}